\documentclass[twocolumn,aps,prd,showpacs]{revtex4}
\usepackage{graphicx}
\usepackage{amsmath}
\usepackage[all]{xy}
\usepackage{amssymb}
\newcommand{\be}{\begin{equation}}
\newcommand{\ee}{\end{equation}}
\newcommand{\ben}{\begin{eqnarray}}
\newcommand{\een}{\end{eqnarray}}
\newcommand{\bes}{\begin{subequations}}
\newcommand{\ees}{\end{subequations}}
\newcommand{\bb}{\bibitem}

\newcommand{\wt}{\widetilde}

\begin{document}
\title{Defect Structures in Lorentz and CPT Violating Scenarios}
\author{M.N. Barreto,$^a$ D. Bazeia,$^a$ and R. Menezes$^{ab}$}
\affiliation{$^a$Departamento de F\'\i sica, Universidade Federal da Para\'\i ba, 58051-970 Jo\~ao Pessoa, Para\'\i ba, Brazil\\
$^b$Centro de F\'\i sica e Departamento de F\'\i sica, Universidade do Porto, Porto, Portugal}
\date{\today}

\begin{abstract}
We investigate models described by real scalar fields, searching for defect structures in the presence
of interactions which explicitly violate Lorentz and CPT symmetries. We first deal with a single field, and we investigate a class of models which supports traveling waves that violate Lorentz invariance. This scenario is then generalized to the case of two (or more) real scalar fields. In the case of two fields, in particular, we introduce another class of models, which supports topological structures that attain a Bogomol'nyi bound, although violating both Lorentz and CPT symmetries. An example is considered, for which we construct the Bogomol'nyi bound and find some explicit solutions. We show that violation of both Lorentz and CPT symmetries induces the appearance of an asymmetry between defects and anti-defects, including the presence of linearly stable solutions with negative energy density in their outer side.
\end{abstract}

\maketitle
\section{Introduction}

The possibility of breaking Lorentz and CPT symmetries has been considered in several different contexts; see, e.g.,
Refs.~{\cite{CFJ,CKL,cordas}}. In \cite{CFJ} the authors modify the usual Maxwell dynamics with the inclusion of a Chern-Simons-like term that violates both Lorentz and CPT symmetries. Other investigations with the addition of contributions that violate Lorentz and CPT symmetries have been done both at low energies, in the standard model \cite{CKL}, and at higher energies, in string-like models \cite{cordas}. For models dealing with CPT and Lorentz violating extensions of the standard model, sometimes one modifies the scalar Higgs sector, and this gives room for defect structures of more general profile, which may play important role to describe phase transitions in the earlier universe, due to spontaneous symmetry breaking.

Defects like domain walls, cosmic strings, monopoles and others have been studied in several different aspects \cite{Inicio}, with applications to Cosmology \cite{Vil} and Condensed Matter \cite{cmat}. In particular, kinks are topological defects which in general connect distinct isolated minima in models that develop spontaneous breaking of some discrete symmetry. They appear in two-dimensional space-time, and can be embedded in the four-dimensional space-time, to generate bidimensional structures named domain walls. The role of such defects as seeds for the formation of non-topological structures is interesting \cite{lee} and has led to several investigations, with the change of the discrete symmetry to an approximate symmetry \cite{as}, and also when the symmetry is biased to make domains of distinct but degenerate vacua spring unequally \cite{b}.
In two-field models, topological defects may generate other interesting structures, such as defect inside defect \cite{did}, and junctions of defects \cite{junction}, and may be of interest in applications concerning conformational structure of polymers and polymer-like chains \cite{polymer}. They may also induce interesting effects on other fields; for instance, the behavior of fermions in the background of kink-like structures is known to have  very significant results \cite{jr}, and could perhaps be re-examined within the Lorentz-violating scenario. 

In this work we study models which combine the two issues, that is, we investigate kink-like structures in scenarios where Lorentz and CPT symmetries may be broken. Our main motivation is related to braneworld, specifically to the Randall-Sundrum scenario \cite{brane}, because we may follow the lines of \cite{bfg} and use the Lorentz-violating model described in Sec.~{\ref{sec:model}} in the context of warped geometry with a single extra dimension. Another motivation is to bring some very well-known results for defect structures in models described by real scalar fields to this new scenario, where Lorentz and CPT symmetries do not play the standard role. In a recent work \cite{ludo}, kinks were investigated in a model which breaks Lorentz symmetry with the explicit inclusion into the Lagrange density of Lorentz non-invariant higher-order derivative contribution. Our route is different, since we will study Lorentz and CPT breaking without introducing higher derivatives. 

To do this, we follow Ref.~{\cite{CKL}}, in which the new terms arise as modifications in the Higgs sector of the standard model. In the light of the recent understanding of equivalence between non-commutative field theory and Lorentz-violanting extensions involving ordinary fields \cite{chkl}, the present work is also of interest to non-commutative solitons \cite{ns}, which has been investigated for a variety of reasons, including self-consistent deformation of the highly constrained structure of local quantum field theory, and the breaking of locality at short distances, which is of direct interest to quantum gravity. Also, the appearing of non-commutativity in field theory in a limit of string theory \cite{string} provides fresh interest to the subject, in particular on D-branes, specially as non-commutative solitons of tachyon fields of open string theory \cite{t}. 

Our investigations consider static solutions in one spatial dimension. Thus, the static solutions that we consider cannot see effects of non-commutativity. However, we can use the point of view of Ref.~{\cite{vy}} to investigate how stability modifies the bound states of the model
for non-commutative space-time. Moreover, our investigations is also of interest to the non-commutative aspects introduced in Ref.~{\cite{kkl}},
which investigates kinks and domain walls for non-commutative field theory, directly connected to the tachyon action for unstable brane in open strings; see the recent revision of Ref.~{\cite{s}} for a variety of motivations on tachyon dynamics.

We organize our work as follows: in the next Sect.~{II} we consider models described by one and by two real scalar fields. There we realize that two-field models lead to richer possibilities, and we show how to extend the Bogomol'nyi bound to the Lorentz and CPT breaking scenario. In Sect.~{III} we investigate an explicit model of two real scalar fields, which can be seen as an extension of a former model, first investigated in \cite{Bazeia}, which has been used in several other contexts, for instance in Refs.~{\cite{did,junction,polymer,Bazeia1}}, engendering broader interest. As we will show, the breaking of both
Lorentz and CPT symmetries gives rise to an asymmetry between defects and anti-defects, including the presence of
linearly stable solutions that support regions of negative energy density.

We end this work in Sect.~{IV}, where we include our comments and conclusions, pointing some possible extensions of this work.

\section{Scalar field models}

In this work we investigate defect structures described by scalar fields in models which break Lorentz and CPT symmetries explicitly.

We start with the simplest case, which describes a single real scalar field. In this case, we study models
where only the Lorentz symmetry is broken. Next, we deal with two scalar fields, and there we
investigate models which break both Lorentz and CPT symmetries.

\subsection{One field}

We start with the model
\be\label{1f}
{\cal L}=\frac12\partial_\mu\phi\,\partial^\mu\phi+
\frac12\kappa^{\mu\nu}\partial_\mu\phi\partial_\nu\phi-V,
\ee
where $V=V(\phi)$ is the potential, which controls the way the field self-interacts. We are working in
$(1,1)$ spacetime dimensions, and the metric is diag$(g_{\mu\nu})=(1,-1),$ with $\kappa^{\mu\nu}$ being a constant tensor, given by 
\be\label{kmunu}
\kappa^{\mu\nu}=\left(\begin{array}{cc}
	\beta & \alpha \\ 
	\alpha & \beta 
\end{array} \right),
\ee
where $\alpha$ and $\beta$ are real parameters. See Ref.~{\cite{KL}} for other details.
For simplicity, we take $\beta=0$ for the explicit calculations that follow. 

The equation of motion is
\be\label{1fem}
\ddot\phi-\phi^{\prime\prime}+2\alpha\dot\phi^\prime+\frac{dV}{d\phi}=0,
\ee
where $\dot\phi=\partial\phi/\partial t$ and $\phi^\prime=\partial\phi/\partial x,$ etc.
For static field we get
\be\label{sf}
\phi^{\prime\prime}=\frac{dV}{d\phi}.
\ee
This is the same equation one gets in the standard situation. Thus, static solutions violate neither
Lorentz nor CPT symmetries. However, for time-dependent field, we search for traveling waves and now
the equation of motion may have solutions which violate Lorentz and CPT symmetries.

Although our model violates Lorentz symmetry, we can still search for traveling 
waves in the form $\phi=\phi(u),$ where $u=\gamma(x-vt),$ but now $\gamma=\gamma(v,\alpha)$ may not have the usual form. We use this into eq.~(\ref{1fem}) to get to
\be\label{tf}
\frac{d^2\phi}{du^2}=\frac{dV}{d\phi},
\ee
if one sets $\gamma=1/\sqrt{1-v^2+2\alpha v}.$ This is a general result: it shows that for any static field $\phi_s(x)$ [topological (kinklike) or nontopological (lumplike)] which solves eq.~(\ref{sf}), there is a traveling wave of the form
\be
\phi(u)=\phi_s(u),
\ee 
which solve eq.~(\ref{tf}). The traveling wave has the form
of a static solution, and it travels with constant velocity $v,$ with width $w= w_0/\gamma,$ for
$w_0$ being the width of the static solution. The velocity is restricted to the interval
$v\in(-\sqrt{1+\alpha^2}+\alpha, \sqrt{1+\alpha^2}+\alpha).$ We notice that the limit $\alpha\to0$ leads to the standard situation, with $\gamma=\gamma(v,0)=1/\sqrt{1-v^2},$ and $v\in(-1,1).$ We also notice that for $\alpha$ very small we get $v\in(-1+\alpha,1+\alpha),$ which shifts by $\alpha$ the standard velocity interval.

We consider the model (\ref{1f}) in the absence of potential; this case was recently considered in
Ref.~{\cite{amw}}, with other motivations. The massless excitations now give $w^2-k^2-2\alpha wk=0,$ which implies that the velocities should obey $v_{\pm}=\pm\sqrt{1+\alpha^2}+\alpha.$ They travel with different velocities in the forward and backward directions, showing that the model engenders birefringence. The inclusion of the potential will make the excitations massive, with velocity bounded by the two massless values. This gives an alternative way to understand the bounds in the velocity of the traveling waves that we have just obtained.

The parameter $\alpha$ induces an asymmetry for traveling waves with positive and negative velocities, breaking Lorentz invariance. We also see that the time-dependent solutions violate both parity and time reversal, although they are symmetric under PT. Thus, they do not violate CPT, because the scalar field is even under charge conjugation.

We calculate $\theta^{\mu\nu}$ to get
the four entries: they are densities which represent energy $\theta^{00}$, energy flux $\theta^{10},$ momentum $\theta^{01},$ and pressure $\theta^{11}$. They are given by
\begin{subequations}\label{tmunu}
\ben
\theta^{00}&=&\frac12\dot\phi^2+\frac12\phi^{\prime2}+V,
\\
\theta^{01}&=&-\dot\phi\phi^\prime-\alpha\phi^{\prime2},
\\
\theta^{10}&=&-\dot\phi\phi^\prime+\alpha\dot\phi^2,
\\
\theta^{11}&=&\frac12\dot\phi^2+\frac12\phi^{\prime2}-V.
\een
\end{subequations}
We notice that both equations $\partial_0\theta^{00}+\partial_1\theta^{10}=0$ and $\partial_0\theta^{01}+\partial_1\theta^{11}=0$ work on shell. The fact that $\theta^{01}\neq\theta^{10}$ indicates violation of Lorentz symmetry.

The energy for traveling waves can be written in the form
\be
\frac{E_v}{E_0}=\gamma\;(1+\alpha v),
\ee
where $E_0$ stands for the energy of the static solutions. We calculate the energy ratio for solutions with opposite velocities to get 
\be\label{e+-}
\frac{E_v}{E_{-v}}= \frac{1+\alpha v}{1-\alpha v}\sqrt{\frac{1-v^2-2\alpha v}{1-v^2+2\alpha v}},
\ee
which is asymmetric, thus violating Lorentz symmetry. We notice that $E_v<E_{-v}$ for $\alpha v>0.$

This is the general scenario for kinks and lumps in models of the form given by eq.~(\ref{1f}). The traveling waves are even under CPT, but they violate Lorentz symmetry.

We illustrate this case with the $\phi^4$ model. It is described by the potential $V(\phi)=(1/2)(1-\phi^2)^2,$ where we are using dimensionless field and coordinates.
The static kink has the form $\phi_s(x)=\tanh{x}.$ It has unit width, and we have chosen $x=0$ as the center of the solution. The corresponding traveling wave is given by $\phi(u)=\tanh\gamma(x-vt),$
which has width $1/\gamma.$ 

We can widen the above investigations using some recent results on deformed defects \cite{dd}.
For the model (\ref{1f}), if one modifies the potential according to 
\be
V(\phi)\to{U}(\varphi)=V(\phi\to f(\varphi))/f^{\prime2}(\varphi),
\ee
where $f=f(\varphi)$ is the deformation function, we can obtain static solution for the modified model in terms of static solution of the starting model. That is, if $\phi_s(x)$ is solution
for the potential $V(\phi),$ then 
\be
\varphi_s(x)=f^{-1}(\phi_s(x)),
\ee
is solution for the modified model with potential $U(\varphi).$ Evidently, the presence
of the Lorentz breaking term in the model (\ref{1f}) does not modify this result, which
shows that the deformation prescriptions introduced in Refs.~{\cite{dd}} are very naturally extended to traveling waves in the above Lorentz violating scenario.

Before going deeper into Lorentz-violating investigations, some words of caution seem to be necessary. It is important to notice that for the model (\ref{1f}) with $\kappa^{\mu\nu}$ given by (\ref{kmunu}), we can redefine field and coordinates in order to eliminate Lorentz violation \cite{km}. This shows that this model is fake Lorentz-violating theory, but we have decided to make the above investigations because it illustrates with simple terms how Lorentz-violating ingredients enter the game for kinks and lumps in $(1,1)$ space-time dimensions. Evidently, the procedure suggested to eliminate Lorentz violation indicates that we can extend the energy-momentum tensor (\ref{tmunu}) in order to make it symmetric and conserved, thus eliminating the presence of Lorentz violation. However, this procedure to eliminate Lorentz violation may not work when we couple the model with more sophisticated fields.

Another issue concerns the need to make the classical solutions time-dependent to make them feel the presence of Lorentz violation. This fact reminds us very much of the investigations done in Ref.~{\cite{vy}}, in which non-commutativity is only seen by the fluctuations around classical static kinks in $(1,1)$ non-commutative space-time. This point will be further explored in a forthcoming investigation, in which we deal with stability of the  Lorentz-violating solutions that appear in this work.

\subsection{Two fields}

We now turn attention to two-field models. Firstly, we consider the class of models
\ben\label{2f1}
{\cal L}&=&\frac12\partial_\mu\phi\partial^\mu\phi+
\frac12\kappa^{\mu\nu}\partial_\mu\phi\partial_\nu\phi+
\nonumber
\\
&&\frac12\partial_\mu\chi\partial^\mu\chi+
\frac12\kappa^{\mu\nu}\partial_\mu\chi\partial_\nu\chi-V(\phi,\chi).
\een
This class of models can be seen as an extension for two fields of the class introduced in the former Sect.~{II.A}. Thus, it also suffers from the same problem of being fake Lorentz-violating theory \cite{km}, but we explore some peculiarities before introducing a genuine Lorentz-violation family of models. Our point is that these models may be seen as effective portions of some more sophisticated models, involving coupling with other more complex fields. 

The equations of motion are given by
\bes
\ben
\ddot\phi-\phi^{\prime\prime}+2\alpha\dot\phi^\prime+\frac{\partial V}{\partial\phi}=0,
\\
\ddot\chi-\chi^{\prime\prime}+2\alpha\dot\chi^\prime+\frac{\partial V}{\partial\chi}=0.
\een
\ees
Thus, for static solutions we get
\bes
\ben
\phi^{\prime\prime}&=&\frac{\partial V}{\partial\phi},
\\
\chi^{\prime\prime}&=&\frac{\partial V}{\partial\chi},
\een
\ees
which do not depend on $\alpha,$ and so they correspond to standard models. The case
\be\label{pot2f}
V(\phi,\chi)=\frac12W_\phi^2+\frac12W_\chi^2,
\ee
where $W_\phi=\partial W/\partial\phi$ and $W_\chi=\partial W/\partial\chi,$ leads to models of the form considered in Refs.~{\cite{Bazeia,Bazeia1}} and in other works.

We consider traveling waves in the form $\phi=\phi(u)$ and $\chi=\chi(u)$ with $u=\gamma(x-vt),$ as before. The equations of motion change to
\bes
\ben
\frac{d^2\phi}{du^2}&=&\frac{\partial V}{\partial\phi},
\\
\frac{d^2\chi}{du^2}&=&\frac{\partial V}{\partial\chi},
\een
\ees
where we have set $\gamma=1/\sqrt{1-v^2+2\alpha v}.$ For this reason, if the model supports
static solutions $\phi_s(x)$ and $\chi_s(x),$ it also supports traveling waves in the form
\be
\phi=\phi_s(u), \;\;\;\;\;\chi=\chi_s(u),
\ee
which travels with constant velocity $v,$ and with width $w=w_0/\gamma,$ as before.

This class of models is similar to the former one, and it may support traveling waves which preserve CPT, although they violate Lorentz symmetry. We notice that extensions to a set of $N$ real scalar fields works straightforwardly.

Another class of models can be considered. In this case the Lagrange density has the form
\be
{\cal L}=\frac12\partial_\mu\phi\,\partial^\mu\phi+\frac12\partial_\mu\chi\partial^\mu\chi+
\kappa^\mu\phi\partial_\mu\chi-V(\phi,\chi).
\ee
The presence of the vector $\kappa^\mu=(a, b),$ $a$ and $b$ being real parameters, leads to
both Lorentz and CPT violation; see Ref.~{\cite{CKL}} for other details. The model may support kinks and lumps, if the potential $V=V(\phi,\chi)$ is chosen properly. This class of models may support defect structures which violate both Lorentz and CPT symmetries, leading to richer scenarios.
In particular, we are now dealing with a genuine Lorentz-violating family of models, since it is not possible to remove the Lorentz-violating 
$\kappa$-dependent term from the theory anymore; see \cite{CKL} and, in particular \cite{K}, in connection with a varying coupling.
 
For the model at hand, the equations of motion have the form
\begin{subequations}
\ben
\partial_\mu\partial^\mu\phi-\kappa^\mu\partial_\mu\chi+\frac{\partial V}{\partial\phi}=0, \label{eqmo1}
\\
\partial_\mu\partial^\mu\chi+\kappa^\mu\partial_\mu\phi+\frac{\partial V}{\partial\chi}=0. \label{eqmo2}
\een
\end{subequations}

The energy-momentum tensor has the four entries:
\begin{subequations}
\ben
\theta^{00}&=&\frac12(\dot\phi^2+\dot\chi^2+\phi^{\prime2}+\chi^{\prime2})-b\,\phi\chi^\prime+V,
\\
\theta^{10}&=&-\phi^{\prime}\dot\phi-\chi^{\prime}\dot\chi+b\phi\dot\chi,
\\
\theta^{01}&=&-\phi^\prime\dot\phi-\chi^\prime\dot\chi-a\phi\chi^\prime,
\\
\theta^{11}&=&\frac12(\dot\phi^2+\dot\chi^2+\phi^{\prime2}+\chi^{\prime2})+a\,\phi\dot\chi-V.
\een
\end{subequations}
We notice that the equations $\partial_\mu\theta^{\mu\nu}=0$ work on shell. Also, $\theta^{01}\neq\theta^{10}$ shows that the model volates Lorentz symmetry. In this case, it is not possible to improve the energy-momentum tensor to make it symmetric and conserved; this is a true manifestation of Lorentz violation for this family of models \cite{CKL,K}. 

For static fields, that is, for field configurations that only depend on the space coordinate $x,$ the equations of motion become
\begin{subequations}\label{2fsf}
\ben
\phi^{\prime\prime}+b\,\chi^\prime &=&\frac{\partial V}{\partial\phi},\\
\chi^{\prime\prime}-b\,\phi^\prime &=&\frac{\partial V}{\partial\chi}.
\een
\end{subequations}
These equations do not depend on $a;$ thus, if one chooses $b$ equal to zero, the static solutions are not affected by Lorentz and CPT symmetries. However, they may be affected by the motion of traveling waves, as we have already shown in the former case.

For nonzero $b,$ we see that the above equations violate both Lorentz and CPT symmetries. They do not respect parity transformation, although they are even under T and C. The absence of parity symmetry breaks the kink $\leftrightarrow$ antikink exchange scenario, which in general appears in models that do not violate parity. However, we notice that the substitutions $x\to-x$ and $b\to-b$ do not change the equations of motion (\ref{2fsf}) for static fields, if the potential is even under $b\to-b.$ In this case, kinks for the model with $b$ positive would become antikinks for the model with $b$ negative. 

The presence of $b$ in the equations of motion and energy density changes the standard scenario. To attain a Bogomol'nyi bound \cite{BPS}
we modify the potential in Eq.~(\ref{pot2f}). We consider a new class of models, identified by
\be\label{pot22}
V_s(\phi,\chi)=\frac12\,(W_\phi+s_1\chi)^2+\frac12\,(W_\chi+s_2\phi)^2,
\ee
where $W=W(\phi,\chi)$ is a smooth function of the two fields, with $s_1$ and $s_2$ being real constants, which obey $s_2-s_1=b$. This potential is an extension of the potential considered in \cite{Bazeia}; it gets to its original form in the limit $b\to0.$
This modification is introduced to attain a Bogomol'nyi bound \cite{BPS}, but it changes the way the fields interact, since the potential now depends explicitly on $b,$ the parameter which breaks Lorentz
and CPT symmetries.

This class of models can be further investigated for the presence of topological solutions. We consider static fields, $\phi=\phi(x)$ and $\chi=\chi(x)$. We write the energy density for static solutions in the form
\be\label{energy}
\theta^{00}\!=\!\frac{d{W}}{dx}+\frac12\left(\phi^\prime\!-\!{W}_\phi\!-\!s_1\chi\right)^2\!+
\frac12\left(\chi^\prime\!-\!{W}_\chi\!-\!s_2\phi\right)^2.
\ee
The energy is minimized to the value $E^{ij}=\Delta{W}_{ij},$ with $\Delta{W}_{ij}={W}_i-{W}_j,$ for ${W}_i={W}(\bar\phi_i,\bar\chi_i),$ and $v_i=(\bar\phi_i,\bar\chi_i)$ being a minimum of the potential, obeying $V(\bar\phi_i,\bar\chi_i)=0.$ This bound is attained for field configurations which obey the first-order equations 
\begin{subequations}\label{foeq1}
\ben
\phi^\prime&=&{W}_\phi+s_1\chi, \\
\chi^\prime&=&{W}_\chi+s_2\phi, 
\een
\end{subequations}
with the boundary conditions: the pair $(\phi,\chi)$ goes to $(\bar\phi_i,\bar\chi_i)$ for
$x\to\infty,$ and to $(\bar\phi_j,\bar\chi_j)$ for $x\to-\infty.$ This is the Bogomol'nyi bound \cite{BPS}, now extended to the above class of models, which violate Lorentz and CPT symmetries.

We can see that solutions of the above first-order equations solve the equations of motion. Also, despite the modification in the model, the static solutions satisfy
\be
\frac12\phi^{\prime2}+\frac12\chi^{\prime2}=V,
\ee
which shows that the gradient and potential portions of the energy contribute equally.

We remark that since $b$ changes the energy density in eq.~(20a), the form (\ref{energy}) is only obtained when we consider the potential in the specific form (\ref{pot22}), with $s_2-s_1=b$. We compare this with the case which preserves Lorentz and CPT symmetries to see that the Bogomol'nyi bound requires the inclusions of new terms into the potential.

The asymmetry that appears for $b\neq0$ may contribute to destabilize the defect solutions. However,
we can show that solutions to the above first-order equations are linearly stable. The calculation
follows the standard route \cite{BS}. The full investigation will be done in another work, and here we show the main steps of the calculation. This investigation is important, because we will show below that there are models which support kinks of unusual profile.
We introduce general fluctuations for the two fields in the form: $\phi(x,t)=\phi(x)+\eta(x,t)$ and $\chi(x,t)=\chi(x)+\xi(x,t).$ We use these fields in the equations of motion to get to the Schr\"odinger-like equation, $H\Psi_n(x)=\omega_n^2\Psi_n(x),$ where $\Psi_n(x)$ is a two-component wave function and the Hamiltonian has the form
\be
H=-\frac{d^2}{dx ^2}-ib\sigma_2\frac{d}{dx}+U,
\ee
where $\sigma_2$ is a Pauli matrix and
\ben
U=\left(\begin{array}{ccc}
{\partial^2 V_s}/{\partial\phi^2} &{\partial^2 V_s}/{\partial\phi\partial\chi} \\
{\partial^2 V_s}/{\partial\chi\partial\phi} & {\partial^2V_s}/{\partial\chi^2}
\end{array} \right).
\een
We use $V_s$ as in Eq.~(\ref{pot22}) to write $H=S^\dag S,$ where $S$ is the first-order operator
\be
S=-\frac{d}{dx}+\left(\begin{array}{lll}
{\;\;\;\;W_{\phi\phi}} & {W_{\phi\chi}+s_1} \\
{W_{\chi\phi}+s_2} & {\;\;\;\;W_{\chi\chi}}
\end{array} \right).
\ee
This shows that $H$ is non-negative, and so the corresponding eigenvalues must obey $w_n^2\geq0.$ This result is general; it extends the result of Ref.~{\cite{BS}} to the above model, and it shows that the solutions of the first-order equations (\ref{foeq1}) are linearly stable.

\section{Example}
\label{sec:model}

The last class of models deserves further attention. We
illustrate this case with an example. We consider $s_1=0$ and $s_2=b,$ and the following function \cite{Bazeia}
\be
W(\phi,\chi)=\phi-\frac13\phi^3- r\phi\,\chi^2,
\ee
where $r$ is a real parameter. This gives the potential
\be\label{pot}
V(\phi,\chi)=\frac12(1-\phi^2-r\chi^2)^2+\frac12\left(2r\phi\chi-b\phi\right)^2.
\ee
The model may support several minima, depending on the values of $r$ and $b.$ We consider $r$ and $b$ positive, and $b^2/4r\in(0,1)$ to write
\be
v_{h\pm}=\left(\pm Q, b/2r\right),\,\,\,\,\,
v_{v\pm}=\left(0,\pm\sqrt{1/r}\right),
\ee
where $Q=\sqrt{1-b^2/4r}.$ There are four minima, two horizontally aligned, and two vertically aligned, as the subscripts indicate. The limit $b\to0$ implies $Q\to1,$ bringing
the minima $v_{h\pm}$ back to $(\pm1,0),$ to the $\phi$ axis, as expected \cite{Bazeia}.

There are five topological sectors, for solutions that solve the first-order equations, one with energy or tension $t_1=(4/3)Q^3,$ and four with tensions degenerate to the value $t_2=(2/3)Q^3.$ As one knows, in the absence of Lorentz and CPT violation, the standard situation engenders BPS and anti-BPS solutions, which connect the minima in the two possible senses. However, parity violation breaks this symmetry, excluding one of the two possibilities. In the model under investigation, for instance, in the more energetic sector, there is only one solution, connecting $v_{h-}\to v_{h+}.$ The same for the other sectors, where there are solutions connecting $v_{h-}\to v_{v+},$ $v_{v+}\to v_{h+},$ $v_{h-}\to v_{v-},$ and $v_{v-}\to v_{h+}.$ In Fig.~[1] we illustrate how the orbits appear connecting the minima of the potential. 

\begin{figure}[!ht]
\centering
\includegraphics[width=.35\textwidth]{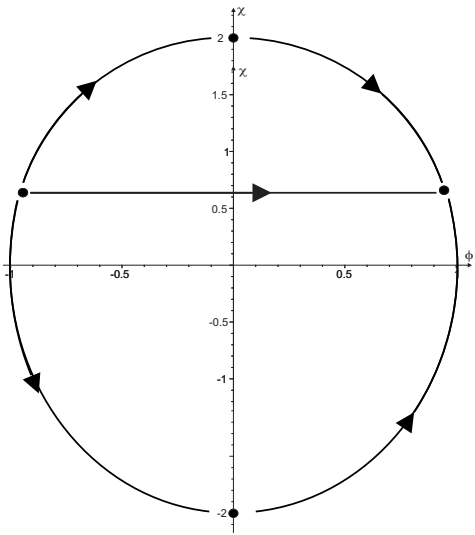}
\caption{The four minima, and some possible orbits for $r=1/4$ and $b=1/3$. The arrows illustrate how the minima are connected for $x$ varying from $-\infty$ to $\infty.$ Parity non-invariance forbids the presence on anti-defects in this case.}
\end{figure}

The model may admit another sector, connecting the minima $v_{v\pm}.$ This sector cannot have solutions that obey the first-order
equations. Although in this case we have been unable to find any explicit solution connecting the two minima asymptotically,
we could verify that the straight line orbit which solves the model for $b=0$ does not exist in the present case, for $b\neq0.$

It is interesting to notice that in a string theory scenario with the above realization of Lorentz and CPT violations, the asymmetry between defects and anti-defects prevent the presence of anti-defects. If this persists in the string theory, it would certainly prevent the presence of open strings ending on a pair brane--anti-brane, and this would certainly change the way tachyon condensation could appear.

For the model under investigation, the first-order equations are 
\begin{subequations}
\label{eq33}
\ben
\phi^\prime&=&1-\phi^2-r \chi^2,\label{eq1}\\
\chi^\prime&=&b\phi-2r\chi\phi.   \label{eq2}
\een
\end{subequations}
It is not hard to see that these equations admit the integrating factor $f(\chi)=1/(\chi-b/2r)^{1+1/r}.$ Thus, we use $\wt\chi=\chi-b/2r$ to write the orbits, for $r\neq1/2$ and $r\neq1,$
\be\label{orb}
\phi^2=\frac{r}{2r-1}{\wt\chi}^{2}+\frac{b}{r-1}{\wt\chi}+
C{\wt\chi}^{\frac{1}{r}}+Q^2,
\ee
where $C$ is an integration constant. The limit $b\to0$ changes this result to the orbits first
obtained in Ref.~{\cite{Izquierdo:2002sz}}.

The specific cases $r=1$ and $r=1/2$ need particular attention. They have orbits given by, respectively,
\begin{subequations}
\ben
\phi^2&=&{\wt\chi}^2+\frac{b}{r}{\wt\chi}\ln{\wt\chi}+C{\wt\chi}+Q^2,
\\
\phi^2&=&C{\wt\chi}^2+{\wt\chi}^2\ln{\wt\chi}-\frac{b}{r}{\wt\chi}+Q^2.
\een
\end{subequations}

We have being unable to solve the first-order equations analytically for $r$ and $C$ arbitrary. For this reason, we have used some specific values for $C$: firstly, we take the limit $C\to\infty,$ to see that in this case the orbit is a straight line segment joining $v_{h+}$ and $v_{h-}$ with $\chi=b/2r.$ This limit reduces the first-order equations (\ref{eq33}) to the single equation $\phi^\prime=Q^2-\phi^2,$ which is solved by
\be \label{sol1}
\phi(x)=Q \tanh(Q x),
\ee
where we are using $x=0$ as the center of the solution. The corresponding energy density is given by $\epsilon=Q^4\,{\rm sech}^4(Q x)$.

Another interesting value for the integration constant is $C=0$. This choice leads to the solutions
\begin{subequations}\label{sol0}
\ben
\phi_{\pm}(x)&=&\frac{Q\sinh(2r Q x)}{\pm B+\cosh(2rQx)},
\\
\chi_{\pm}(x)&=&\frac{b}{2r}\pm\frac{A}{\pm B+\cosh(2rQx)},
\een
\end{subequations}
where we have used $A=(1-r)Q^2K$ and $B=bK/4,$ where
\be\label{cons}
K=\sqrt{\frac{1-2r}{r(1-2r+r^2Q^2)}},
\ee
with $r\in(0,1/2).$

We notice that the limit $b\to0$ changes the solutions (\ref{sol0}) to the simpler
form
\begin{subequations}
\ben
\phi^{0}(x)&=&\tanh(2r x),
\\
\chi^{0}_{\pm}(x)&=&\pm\sqrt{\left(\frac{1}{r}-2\right)}\;{\rm sech}(2r x),
\een
\end{subequations}
which are solutions of the model first investigated in Ref.~\cite{Bazeia}. 
We recall that the above solutions were found with the elliptic orbits
\be
\phi^2+\frac{r}{1-2r}\chi^2=1,
\ee
which are good orbits for $r\in(0,1/2).$ We notice that these orbits are exactly the orbits obtained in eq.~(\ref{orb}) in the limit $b\to0$ for the value $C=0.$

The energy density corresponding to the above solutions can be written as
\be
\theta^{00}=\phi^{\prime2}+\chi^{\prime2}-b\phi\chi^\prime,
\ee
and for the non-trivial solutions with $C=0$ we use Eqs.~(\ref{sol0}) and (\ref{cons}) to obtain
\ben
\theta_{\pm}^{00}(x)&=&\frac{4r^2Q^4}{[B\pm\cosh(2rQx)]^4}\bigg[1+B^2\pm\nonumber
\\
& &\cosh(2rQx)\left(2B+\frac{bA}{2rQ^2}\sinh^2(2rQx)\right)+\nonumber
\\
&&\left(\frac{A^2}{Q^2}+\frac{bAB}{2rQ^2}+B^2\right)\sinh^2(2rQx)\bigg].
\een

The orbits and solutions for $C=0$ are shown in Fig.~[2] and
[3], respectively, and in Fig.~[4] we plot the corresponding energy densities.
These figures are shown for $r=1/4$ and $b=1/3.$ We see that the upper orbit gives
standard defect structures. However, the lower orbit gives unusual defects, making the
topological solution non monotonic, a fact due to the breaking of Lorentz invariance,
which also responds for the presence of regions of negative energy density, as shown in Fig.~[4].
\begin{figure}[!ht]
\centering
\includegraphics[width=.38\textwidth]{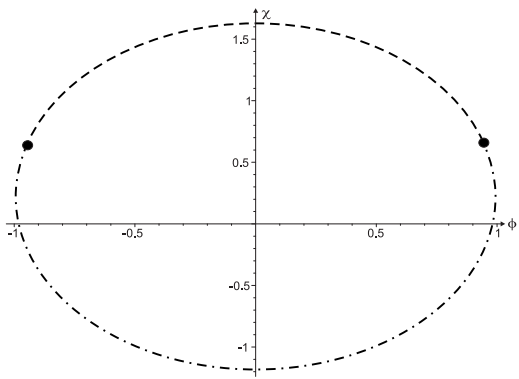}
\caption{Orbits for defect solutions in the sector connecting the minima $v_{h\pm}=(\pm Q,s)$ for $C=0.$ The upper and lower orbits are depicted with dashed and dot-dashed lines, using $r=1/4$ and $b=1/3.$}
\end{figure}
\begin{figure}[!ht]
\centering
\includegraphics[width=.45\textwidth]{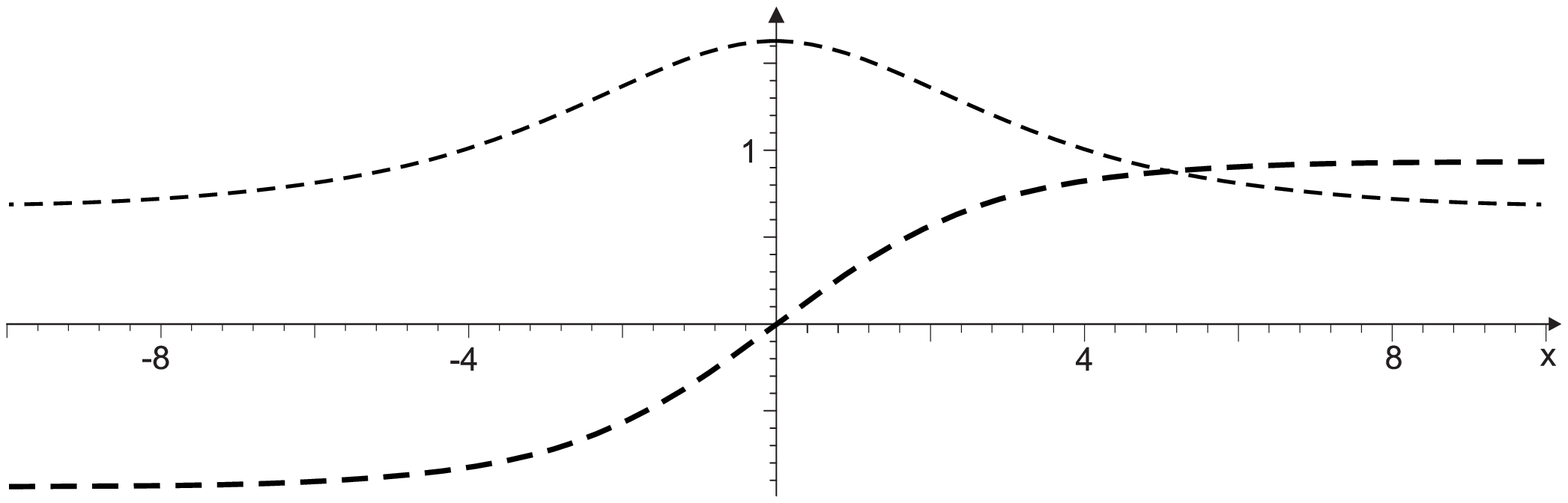}
\vspace{.4cm}
\includegraphics[width=.45\textwidth]{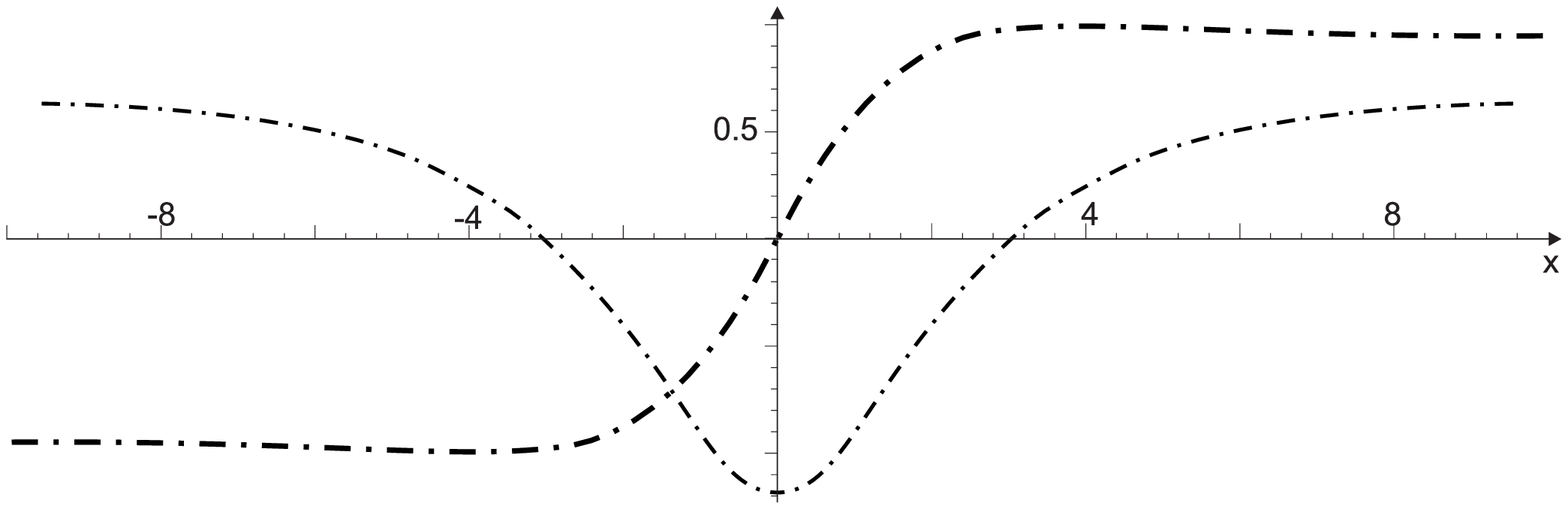}
\caption{Field profile for defect solutions corresponding to upper and lower orbits shown in Fig.~[2], plotted in the upper and lower panels, respectively. We distinguish the two fields with thicker $(\phi)$ and thinner $(\chi)$ lines, and we use $r=1/4$ and $b=1/3.$}
\end{figure}

To introduce specific results, we notice that in the defect solution for lower orbit, the behavior of the $\phi$ field, which ensures the topological profile of the solution, shows two critical points, at the values $x^{\pm}_c=\pm (1/2rQ){\rm arcsech(B)},$ for which $\phi(x^{\pm}_c)=\pm Q/\sqrt{1-B^2\,}.$ For these values, the energy density is given by
\be
\theta_-^{00}(x_c)=-rb^2\frac{(1-2r)(1-2r+r^2Q^2)}{(1-r)^4},
\ee
and it is always negative, for the range of values that we are considering. For $r=1/4$ and $b=1/3$ we get $x^{\pm}_c=\pm 3.8575.$ Although the energy is positive, the energy density is negative in the two regions $|x|\geq 3.0625,$ which include the critical points of $\phi;$ see Fig.~[4]. These regions of negative energy densities form the outer side of the defect, and they disappear in the limit $b\to0,$ in the absence of Lorentz and CPT breaking. The core of the defect changes insignificantly for $b$ small, and so it may entrap another defect in the same way it used to do in the standard situation \cite{did}. The appearance of negative energy density is an unusual behavior, which leads us to think that such solutions are unstable, but we have already show that they are linearly stable in general. We will further investigate stability in another work, to examine how to find stable solutions for specific models which violate both Lorentz and CPT symmetries.

\begin{figure}[!h]
\centering
\includegraphics[width=.45\textwidth]{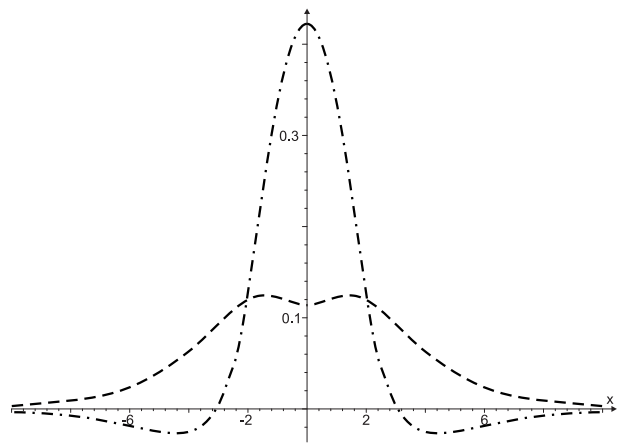}
\caption{Energy density in the case $C=0,$ with $r=1/4$ and $b=1/3.$ The dashed and dot-dashed lines
correspond to solutions for the upper and lower orbits, respectively, as they appear in Fig.~[2].}
\end{figure}

The value $b=1/3$ is not small. Since $b$ measures how the model deviates from
the standard situation, it should be very small. Former studies on bounds in the Higgs sector for extensions of the standard model suggest the order of magnitude of $b.$ The constraint is very tight in more realistic situations \cite{bound}. In our toy model, however, we have used $b=1/3$ to highlight the effects the breaking of Lorentz and CPT symmetries may induce in the defect structures that appear in the model under consideration. Moreover, the present investigations may be of some use in applications to condensed matter -- see, e.g., Refs.{\cite{cmat,polymer}} -- and there violation of Lorentz invariance should have another interpretation. Indeed, in condensed matter we have found interesting investigations \cite{m} in which one deals with very similar solutions, engendering profiles of almost the same type of the kink-like solutions that appear for a non-vanishing $b,$ not that much small. We can also mimic Lorentz-violating models in condensed matter with materials which naturally select preferable directions in space, which can be described with continuum version of the Dzyaloshinkii-Moriya model \cite{DM}.  

\section{Comments and Conclusions}

In this work we have investigated models described by real scalar fields, in
scenarios which violate both the Lorentz and CPT symmetries. We first dealt with models described by a single real field, and there we have shown that the addition of the Lorentz breaking term changes no static sector of the model. However, traveling waves see the Lorentz breaking, and we have constructed the way the traveling waves appear. Moreover, we have extended this result to deformed defects, and to models described by two or more real scalar fields.

In the case of two fields, we have invented another class of models, and we have investigated an explicit example, which generalizes former results to the Lorentz and CPT breaking scenario. These models do not support the usual defect and anti-defect structures
simultaneously, and there are solutions that engender unusual profile, making the energy density negative in the outer side of the defect. The asymmetry for defect and anti-defect that we have found may perhaps be of some use to build string theory scenarios where open strings ending on a brane--anti-brane system are suppressed by CPT violation. 

The present investigations will continue in another work, where we study linear stability of the solutions that we have just found in this paper. There we will show explicitly how to construct stable defect structures which violate Lorentz and CPT symmetries. We will also investigate supersymmetric extensions \cite{rc,bbc,prl} of the above models, to see how the solutions of the first-order equations behave as BPS states.

We believe that the idea that the geodesic motion in moduli space can be used to describe the low energy dynamics of defect structures \cite{manton} may be extended to the present context. Eventually, it may change the scenario constructed in \cite{es} for the standard model, which preserves both
Lorentz and CPT symmetries.

The suggestion that the models here studied may mimic features of more realistic systems, can also be extended to the case of heterotic M theory, following the recent work \cite{ant}, which has investigated the effects of collision of scalar field kinks with boundaries, motivated from its cousin, the five dimensional heterotic M theory. The investigation shows that kink-boundary effects appears as direct application of the moduli space evolution.

Other lines of investigations concern the presence of junctions of defects, in Lorentz and CPT violating scenarios. Work on this is now in progress, in models which follow the lines of Ref.~{\cite{junction}}. We are also exploring similar models, with focus on tachyon kinks, motivated by ideas present in Refs.~{\cite{t,kkl,s,tk}}. Furthermore, the inclusion of fermions is important not only for supersymmetry, but also to allow investigations concerning the behavior of fermions \cite{jr} in the background of these Lorentz-violating kink-like structures. Another issue concerns the use of defect structures in scalar field theory to generate brane in warped geometry with a single extra dimension, as motivated by Ref.~{\cite{brane}}.
Practical possibilities have already been examined in Ref.{\cite{bfg}}, and we are now searching for brane within the present Lorentz-violating
scenario. Evidently, the presence of Lorentz violation requires that we somehow modify the standard scenario, with the addition of extra terms to compensate the asymmetry of the energy-momentum tensor. Similar recent investigation was done in \cite{jp}, where a Chern-Simons modification of General Relativity has been considered, which may help us enlighten the issue.

\vspace{.6cm}

The authors would like to thank F.A. Brito, A.R. Gomes, L. Losano, J.R. Nascimento and V.M. Pereira for discussions, and CAPES, CNPq, PADCT/CNPq, PROCAD/CAPES, and PRONEX/CNPq/FAPESQ for financial support.

\end{document}